# Interplay between near-field radiative coupling and space charge effects in a micro gap thermionic energy converter under fixed heat input

Ehsanur Rahman[1,2]* and Alireza Nojeh[1,2]

[1] Department of Electrical and Computer Engineering, University of British Columbia, Vancouver, BC, V6T 1Z4, Canada

[2] Quantum Matter Institute, University of British Columbia, Vancouver, BC, V6T 1Z4, Canada

*Email: ehsanece@ece.ubc.ca

**Abstract**

We investigate the performance of a micro gap vacuum thermionic energy converter considering the loss mechanisms due to the space charge effect and interelectrode radiative heat transfer. The dependencies of the space charge effect and near-field radiative heat exchange on the interelectrode distance are derived based on established theories. The electrode temperatures are determined by solving the steady-state energy balance equations in a numerical, iterative process and considering a constant energy flux input to the emitter. The resultant behaviour of the different mechanisms of energy flow from the electrodes is studied for a wide range of interelectrode distances, which provides insights into the device operation. The maximum efficiency of the converter is obtained by optimizing the operating voltage and interelectrode distance. Considering the interplay between space charge and near-field radiative heat transfer, an optimal range is determined for the interelectrode distance. The optimal value of the distance and the lower limit of this range are found to be significantly higher than previously reported, where constant electrode temperatures had been assumed.

## 1. Introduction

Static heat-to-electricity converters offer distinct advantages over turbine generators due to their silent operation, lack of moving parts, high specific power, and flexible form factors. These diverse features are particularly attractive for applications which require portability, relatively low maintenance, and long lifetime. A prominent example of static heat-to-electricity generation mechanisms is thermionic conversion, which has significant advantages over other static energy conversion mechanisms such as the thermoelectric and photovoltaic effects. For example, in a photovoltaic device, incident photons with energy higher than the bandgap can be used to generate electron-hole pairs, which subsequently undergo electronic transitions across the bandgap. However, a significant portion of the solar spectrum is wasted since many of the incident photons either do not have enough energy to excite an electron to the conduction band or have excessive energy, which is lost as heat [1,2]. To improve the photovoltaic efficiency, multijunction solar cells have been proposed, where several epitaxially grown layers of different bandgap materials are stacked together to increase the usable portion of the solar spectrum. However, these devices face significant engineering and materials-based challenges [3]. In this regard, a thermionic energy converter (TEC) has a fundamental advantage as it can, in principle, utilize any source of input heat irrespectively of its spectral composition which, in the case of solar energy, can be obtained from a concentrated solar thermal harvesting system. On the other hand, thermoelectric generators suffer from a low figure of merit due to the combined requirements of high electrical and low thermal conductivity [4,5], ultimately limiting the conversion efficiency to around 10% in practice. Unlike the situation in thermoelectric generators, the vacuum gap in a TEC prevents the problem of lattice thermal conduction. Due to these benefits, thermionic conversion has attracted great attention over the years, with significant experimental demonstrations since the 1950s, utilizing heat from diverse sources [6,7]. The Jet Propulsion Laboratory developed TECs under the Solar Energy Thermionics program in the 1960s [8]. Prototypes were demonstrated, such as the JG-2 device which achieved an electrical power output of 114 W and an energy conversion efficiency of 7.0% when heated to a temperature of 1727 °C. In later efforts, The Japan Solar Upper Stage program designed a TEC which was estimated to operate at an efficiency of 23.2% with a power density of 5.7 W/cm$^2$ at an emitter temperature of about 1577 °C [9]. General Atomics proposed the High Power Advanced Low Mass solar thermionics space power system [10,11], and an early prototype achieved a maximum electrical power of 30 W at an emitter temperature of 1397 °C under direct solar heating. Apart from concentrated solar heating, thermionic converters with other sources of heat have been experimentally demonstrated. For example, TOPAZ reactor (a Russian acronym for Thermionic Experiment with Conversion in Active Zone) was developed by the former Soviet Union (USSR) [12–14]. In 1986 and 1987, the USSR launched two reconnaissance satellites into the lower earth orbit powered by its 6-KW TOPAZ thermionic reactor. These examples show that TECs have strong potential for electricity generation, especially with improvements to existing systems. For further development of thermionic energy conversion technology, two issues which need to be addressed are the space charge effect and radiative heat loss in the interelectrode space. The space charge effect is caused by the mutual repulsion of electrons traversing the interelectrode gap and is detrimental to TEC performance. The additional barrier caused by the space charge substantially reduces the electron flux from the emitter to the collector and can thus reduce the output power and the conversion efficiency by orders of magnitude. A common approach to mitigating the space charge effect has been to use a cesium plasma in the interelectrode space. However, the ionization of the cesium vapour requires additional power and auxiliary discharge electrodes, which reduces the efficiency and adds complexity [15,16]. Moreover, the requirement of the continuous supply of cesium ions limits the operational lifetime of such a converter. Another approach to solving the space charge problem is to create an accelerating field in the interelectrode region by inserting a gate electrode, resulting in a triode configuration [17]. A triode structure could be relatively complex due to the additional requirements of a magnetic and electric field setup [17,18]. As an alternative to the above approaches, the space charge problem may be solved by making the interelectrode gap small enough to prevent the build-up of significant electron clouds in that region. Such micro gap TECs may have advantages over their macro gap counterparts (e.g. plasma or triode TECs) due to fewer components and relatively more compact device structure (although they involve more stringent fabrication requirements than macro gap devices), thereby paving the way to building powerplants on a chip. In such a micro gap vacuum TEC, to achieve a practical current and output power density in a reasonable emitter temperature range, the interelectrode distance (which we sometimes also refer to as the gap size or gap width), needs to be of the order of only a few micrometers [18]. Several efforts have been made recently to develop microfabricated TECs [19–21]. There have also been theoretical studies of the space charge effect in TECs using various analytical and numerical techniques such as a self-consistent Poisson-Vlason solution [22], employing genetic algorithm-based optimization [23], and particle-in-cell modeling [24]. However, these works did not consider the gap size dependence of radiative coupling which is important for micro gap TEC performance analysis. This is because when the interelectrode distance becomes comparable to the characteristic wavelength of thermal radiation (given by Wien's displacement law), near-field radiative heat transfer becomes significant [25–28]. In this small gap range, the far-field radiative heat transfer governed by the Stefan-Boltzmann formula is no longer valid; the total radiative heat transfer may increase by orders of magnitude. This means that there is a trade-off between space charge mitigation and minimizing near-field radiative heat transfer, and that the optimal gap for maximum efficiency should be determined considering the interplay between the two. Therefore, it is worth investigating the performance of a vacuum TEC comprising a microscale gap considering different loss processes. A previous theoretical study to investigate the effects of microscale radiative coupling in TEC performance can be found in the literature [29] for the case where the emitter and collector were held at fixed temperatures while the distance between them was varied. This assumption of fixed electrode temperatures resulted in orders of magnitude variation in the amount of input heat energy required to maintain those temperatures for different values of the interelectrode gap. That is a reasonable assumption when the converter is in direct thermal contact with a large, fixed-temperature thermal reservoir, with much greater energy available than the TEC uses. However, in many other practical scenarios, the amount of input power is fixed. For example, in a concentrated solar thermal system, the intensity of the thermal energy available from solar heating strongly depends on the concentration ratio. This concentration ratio depends on the geometric structure of the focusing system [30–32] which, for example, could be a parabolic trough or disc concentrator. These focusing systems are carefully designed for optimal concentration ratios considering various trade-offs and typically are not adjustable once implemented. Considering these constraints on the input heat source, one needs to determine the electrode temperatures (which are dictated by energy balance) dynamically while varying the interelectrode distance.

In the present study, we model a TEC by analyzing the dependence of the electrode temperatures on the interelectrode distance. This provides important insights into the energy exchange channels between the two electrodes, electric current, and power output of TECs. This model also allows us to determine the optimal interelectrode distance and maximum device efficiency, where the interplay between the near-field radiation and the space charge effect is considered. The optimal gap found is more than twice the value reported in [29] (which was for fixed electrode temperatures) for comparable conversion efficiencies. Apart from insights into the physics of various energy exchange mechanisms within the device, the value of the optimal gap is another significant finding from an application perspective considering the great challenges of fabricating micro gap devices, where a twice-larger gap represents much less stringent fabrication requirements.

**2. Simulation Methodology**

**2.1. Energy balance, efficiency, and the iterative model to solve the coupled energy exchange problem in a micro gap thermionic energy converter**

The input and output energy fluxes at the emitter and collector of the TEC are shown in fig. 1(a). The input heat flux, $Q_{In}$, is absorbed by the emitter, which rapidly heats the electrons and the lattice. Some of the electrons gain sufficient energy to overcome the emitter vacuum barrier and are emitted into the interelectrode space. The energy carried by these thermionically emitted electrons is denoted as $Q_T$. Due to the temperature difference between emitter and collector, a part of the incident energy flux, $Q_{Rad}$, is exchanged between the two electrodes radiatively. As will be discussed later, this interelectrode radiative heat transfer consists of contributions from propagating waves and evanescent waves at small gaps. A part of the incident energy is also lost as radiation from the emitter to the ambient and is given by $Q_{Loss} = \varepsilon\sigma(T_E^4 - T_0^4)$, where $\varepsilon$ is the effective thermal emissivity of the emitter, $\sigma$ is the Stefan Boltzmann constant, and $T_E$, $T_0$ are the emitter and ambient temperatures, respectively. In steady state, the input energy flux into the emitter is equal to the sum of the fluxes leaving from the emitter to the collector and the surroundings:

$$Q_{In} = Q_T + Q_{Rad} + Q_{Loss} \quad . \quad (1)$$

A portion of the thermionic energy flux transferred from the emitter to the collector is converted into useful electrical output. If $J$ is the net current density from the emitter to the collector and $V$ is the operating voltage, then the output power density and efficiency of the thermionic converter can be defined, respectively, as

$$P_{TEC} = JV \quad (2.a) \quad \text{and} \quad \eta = JV / Q_{In} \quad . \quad (2.b)$$

The rest of the energy reaching the collector is dissipated as heat and released to a heat sinking mechanism such as a cooling fluid. The heat flow from the collector to the cooling fluid can be written as

$$Q_{Sink} = K_L(T_C - T_0) \quad (3)$$

, where $K_L$ is the heat transfer coefficient between the cooling fluid and the collector and $T_C$ is the collector temperature. A flow chart of the self-consistent iterative algorithm developed in this work to determine the electrode temperatures and different energy exchange channels is shown in fig. 1(b).

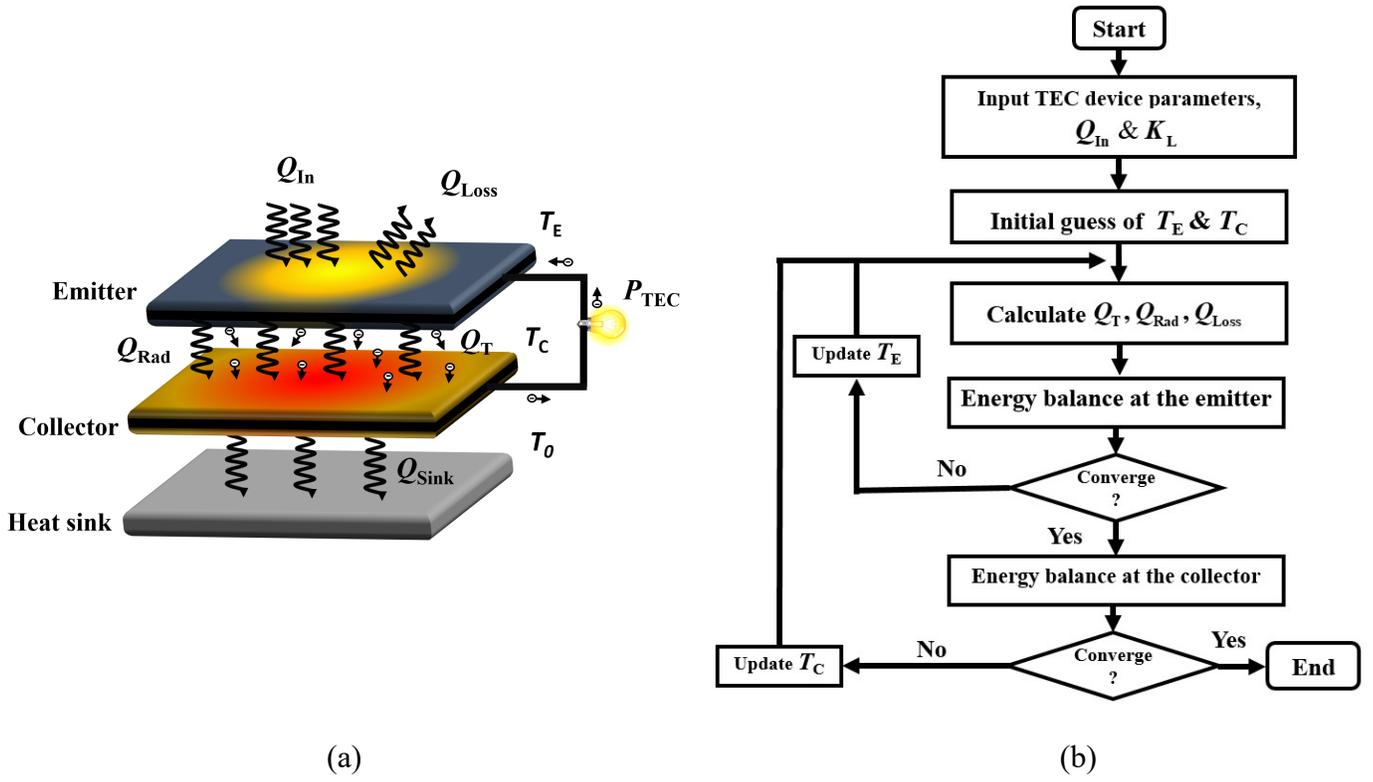

Fig. 1. (a) The input, intermediate, and output energy fluxes in a micro gap thermionic energy converter. (b) A flowchart of the self-consistent numerical iterative model implemented in this work.

The space charge and near-field radiative heat transfer effects are modelled using theories established in the literature. For completeness and ease of access, we provide an overview of the implementation of each of these theories in our model before presenting our results.

## 2.2. The space charge effect in the interelectrode region

The space charge effect in a TEC arises from the coulombic repulsion caused by the electrons in the interelectrode space. The energy diagram of a space charge limited TEC is shown in fig. 2(a). As can be seen, an energy barrier is formed in the gap when the interelectrode electron concentration is very high (which is typically the case for macroscopic gaps). Due to this barrier, only a portion of the electrons overcoming the emitter work function, which have sufficient energy to also surpass this barrier, can reach the collector; lower energy electrons are reflected back to the emitter. This phenomenon significantly reduces the output current density.

The space charge effect is derived assuming that electrons traversing the interelectrode distance are collisionless particles. Under this assumption, the additional barrier, $\varphi(x)$, arising from the mutual repulsion of the electrons in the interelectrode space can be obtained from the Poisson equation,

$$\frac{d^2\varphi}{dx^2} = -e^2 \frac{N(x)}{\varepsilon_0} \quad (4)$$

, where $N(x) = \int_{-\infty}^{+\infty} dv_z \int_{-\infty}^{+\infty} dv_y \int_{-\infty}^{+\infty} dv_x f(x,v)$ is the electron density at the position $x$, $f(x,v)$ is the electron velocity distribution function, and $\varepsilon_0$ is the permittivity of free space. Assuming that the electron velocity distribution is a half Maxwellian at the position of maximum motive, $x_m$, eq. (4) can be rewritten in terms of the dimensionless potential barrier, $\gamma$, and the dimensionless distance, $\xi$, as

$$2\frac{d^2\gamma}{d\xi^2} = e^\gamma [1 \pm erf(\sqrt{\gamma})] \quad (5)$$

, where $\gamma = \frac{\varphi_m - \varphi(x)}{k_B T_E}$, $\xi = \frac{x - x_m}{x_0}$, $x_0 = \sqrt{\frac{\varepsilon_0 k_B T_E}{2e^2 N(x_m)}}$, and $erf(z) = \frac{2}{\sqrt{\pi}} \int_0^z e^{-t^2} dt$ is the error function [18].

In the above definitions, $k_B$ is the Boltzmann constant and $N(x_m)$ is the electron density at the position of maximum motive. The '+' sign applies for $\xi < 0$ and the '-' sign is for $\xi \geq 0$.

Double integrating eq. (5) with appropriate boundary conditions leads to

$$\xi = -\int_0^{\gamma} \frac{dt}{[e^t \pm e^t erf(\sqrt{t}) \mp 2\sqrt{\frac{t}{\pi}} - 1]^{\frac{1}{2}}} \quad (6)$$

, where the upper sign applies for $\xi < 0$ and the lower sign applies for $\xi \geq 0$. We calculated the value of this integral numerically for a wide range of $\gamma$. The value of $\varphi_m$ depends on the operating voltage, which can be summarized for the saturation, space charge and retarding modes of operation [18,33] as

$$\varphi_m = \begin{cases} \varphi_E, & 0 < V < V_S \\ \varphi_E + \gamma_E k_B T_E, & V_S < V < V_C \\ \varphi_C + eV, & V > V_C \end{cases} \quad (7)$$

, where $\gamma_E$ is the value of $\gamma$ at the emitter surface and is given by $\gamma_E = \ln(J_{ES}/J_E)$, in which $J_{ES}$ is the emitter saturation current density and $J_E$ is the emitter current density at voltage $V$. $V_S$ and $V_C$ are the saturation and critical point voltage, respectively, and are defined as follows. When $V = V_S$, the maximum energy barrier occurs just outside the emitter, and all electrons originating from the emitter can reach the collector. This voltage can be expressed as

$$eV_S = \varphi_E - \varphi_C - \gamma_C(\xi_{CS})k_B T_E \quad (8)$$

, where $\gamma_C$ is the value of $\gamma$ at the collector surface and $\xi_{CS} = 9.186 \times 10^5 \sqrt{J_{ES}} d / T_E^{3/4}$ [33]. When $V = V_C$, the maximum energy barrier occurs just in front of the collector and the electrons originating from the emitter need to overcome a decelerating force. The critical point voltage $V_C$ can be defined as

$$eV_C = k_B T_E \ln(A T_E^2 / J_{ER}) - \varphi_C \quad (9)$$

, where $J_{ER}$ is the emitter current density at the critical point [33], which is calculated precisely in the present study using the method described in [34]. For $V < V_S$, the emitter current density is the saturation current density, $J_{ES} = AT_E^2 e^{(-\frac{\varphi_E}{k_B T_E})}$, and for $V > V_C$, it is given by $J_E = AT_E^2 e^{(-\frac{\varphi_C + eV}{k_B T_E})}$ and the device is in retarding mode. For $V_S < V < V_C$, the device is in the space charge regime and the emitter current can be calculated using the method described in [18], which we briefly review here. For a given $\varphi_E$ and $T_E$, we first determine the emitter current density at the saturation point ($J_{ES}$) and critical point ($J_{ER}$). Then, for a particular current density $J_E$ in the space charge region ($J_{ER} < J_E < J_{ES}$), we calculate $\gamma_E$. Using this value of $\gamma_E$, we then calculate $\xi_E$, then $\xi_C$ and finally $\gamma_C$ using the curve resulting from eq. (6). The voltage in the space charge region can then be calculated from $eV = \varphi_E - \varphi_C + (\gamma_E - \gamma_C)k_B T_E$. For the numerical implementation of this method, we follow the algorithm described in [34].

The reverse current from the collector in the saturation, space charge and retarding regions can be defined as

$$J_C = AT_C^2 e^{(-\frac{\varphi_m - eV}{k_B T_C})}. \quad (10)$$

Using the above analysis, the net energy flux carried by the thermionic current from the emitter is given by

$$Q_T = \frac{[(J_E - J_C)\varphi_m + 2k_B(T_E J_E - T_C J_C)]}{e}. \quad (11)$$

The first term in eq. (11) is due to the potential energy and the second term is due to the average thermal energy [35]. A part of this thermionic energy flux is converted to electricity while the rest is deposited in the collector as heat when thermionic electrons are absorbed by it.

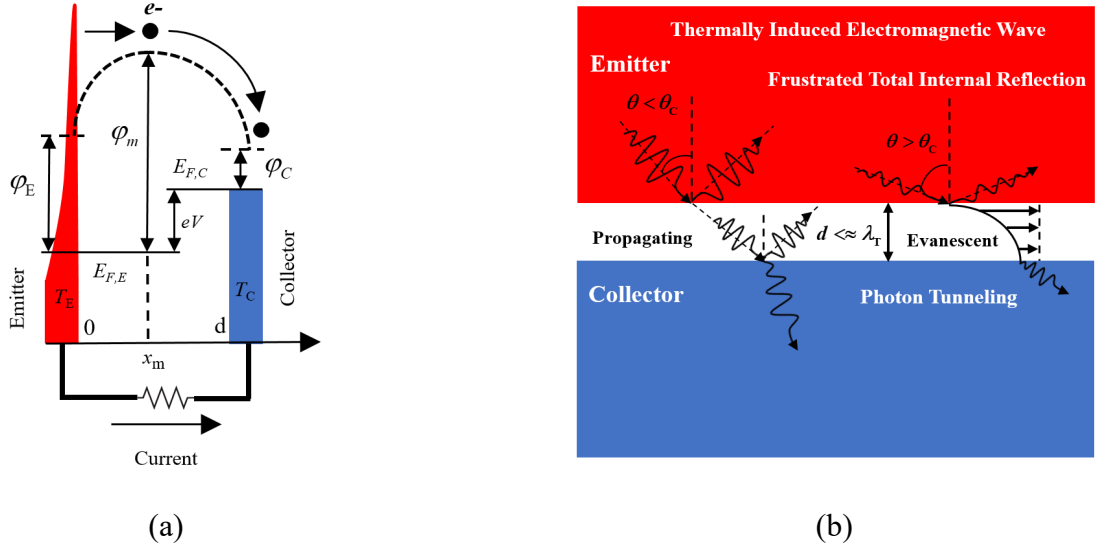

(a)                                                                                     (b)

Fig. 2. (a) The energy diagram of a TEC in the space charge regime. $E_{F,E}$ and $E_{F,C}$ are the fermi levels of the emitter and the collector, respectively. $\varphi_m$ is the maximum motive in the interelectrode space and $e$ is the electron charge. $\varphi_E$ and $\varphi_C$ are the work functions of the emitter and the collector, respectively. $T_E$ and $T_C$ are the temperatures of the emitter and the collector, respectively. $V$ is the voltage difference between the two electrodes. $x_m$ is the position of the maximum motive and d is the interelectrode gap width. (b) Enhanced radiative exchange between the emitter and collector (separated by a vacuum gap) due to near-field coupling of evanescent waves (also known as photon tunnelling). $\theta_C$ is the critical angle of incidence at the emitter-vacuum interface, and $\lambda_T$ is the characteristic wavelength of thermal radiation, given by Wien's displacement law.

## 2.3. Radiative heat transfer between emitter and collector

When the interelectrode distance in a thermionic device is large, radiative heat exchange between the electrodes is due to the far-field propagating waves and is given by the Stefan-Boltzmann law. However, mitigating the space charge effect requires the electrodes to be placed at a distance of a few micrometers or less. This is on the order of the characteristic wavelength of thermal radiation from the emitter, which is given by Wien's displacement law as $\lambda_T = 2.9 x 10^{-3} / T_E$, and so the emitter and collector surfaces are coupled by evanescent waves (fig. 2(b)). The coupling of these evanescent waves significantly enhances the radiative heat transfer in the near-field regime. The energy transfer including both propagating and evanescent components can be modelled using fluctuational electrodynamics [25,28] as

$$Q_{\text{prop}} = \frac{1}{\pi^2} \int_0^\infty d\omega \, [\Theta_E(\omega, T_E) - \Theta_C(\omega, T_C)] \times \int_0^{\omega/c} S_{\text{prop}}(\omega, \beta, \varepsilon_E, \varepsilon_C) \, d\beta \quad (12)$$

and

$$Q_{\text{evan}} = \frac{1}{\pi^2} \int_0^\infty d\omega \, [\Theta_E(\omega, T_E) - \Theta_C(\omega, T_C)] \times \int_{\omega/c}^\infty S_{\text{evan}}(\omega, \beta, \varepsilon_E, \varepsilon_C) \, d\beta \quad (13)$$

, where $Q_{\text{prop}}$ and $Q_{\text{evan}}$ are the heat radiation fluxes due to propagating and evanescent waves photons, respectively. In the above equations, $\Theta(\omega, T_j) = \hbar\omega / (e^{\frac{\hbar\omega}{k_B T_j}} - 1)$ is the mean energy of a Planck oscillator at angular frequency $\omega$, $T_j$ is the temperature of electrode j (emitter or collector), $\varepsilon_j$ is the complex dielectric permittivity of medium j, $\beta$ is the wavevector component parallel to the interface, $c$ is the speed of light, and $\hbar$ is the reduced Planck constant. $S_{\text{prop}}$ and $S_{\text{evan}}$ are the coupling coefficients for the propagating and evanescent waves, respectively. For heat exchange between two semi-infinite plates, these two terms are given by [25,26]

$$S_{\text{prop}}(\omega, \beta, \varepsilon_E, \varepsilon_C) = \frac{\beta(1 - |r_{0E}^s|^2)(1 - |r_{0C}^s|^2)}{4|1 - r_{0E}^s r_{0C}^s e^{i2dk_0}|^2} + \frac{\beta(1 - |r_{0E}^p|^2)(1 - |r_{0C}^p|^2)}{4|1 - r_{0E}^p r_{0C}^p e^{i2dk_0}|^2}, \quad \beta < \frac{\omega}{c} \quad (14)$$

and

$$S_{\text{evan}}(\omega, \beta, \varepsilon_E, \varepsilon_C) = \frac{\text{Im}(r_{0E}^s)\text{Im}(r_{0C}^s)\beta e^{-2d\,\text{Im}(k_0)}}{|1 - r_{0E}^s r_{0C}^s e^{-2d\,\text{Im}(k_0)}|^2} + \frac{\text{Im}(r_{0E}^p)\text{Im}(r_{0C}^p)\beta e^{-2d\,\text{Im}(k_0)}}{|1 - r_{0E}^p r_{0C}^p e^{-2d\,\text{Im}(k_0)}|^2}, \quad \beta > \frac{\omega}{c}. \quad (15)$$

In eqs. (14) and (15), the first term of the right-hand side refers to the contribution of s polarization and the second term refers to the contribution of p polarization. $r_{0j}^s$ and $r_{0j}^p$ are the Fresnel reflection coefficients for s and p polarizations, respectively, at the interface between vacuum and electrode j (emitter or collector) [28] as given by

$$r_{0j}^s = (k_0 - k_j)/(k_0 + k_j) \quad (16.a) \quad \text{and} \quad r_{0j}^p = (\varepsilon_j k_0 - k_j)/(\varepsilon_j k_0 + k_j) \quad (16.b)$$

, where $k_0$ and $k_j$ are the perpendicular wavevector components in the vacuum and medium j, respectively and are given by

$$k_0 = \sqrt{(\omega/c)^2 - \beta^2} \quad (17.a) \quad \text{and} \quad k_j = \sqrt{\varepsilon_j (\omega/c)^2 - \beta^2}. \quad (17.b)$$

We are inspired by the Drude model for tungsten [36] to describe the dielectric permittivity of the electrodes as

$$\varepsilon_j(\omega) = 1 - \frac{\sigma_0/\tau_j}{\varepsilon_0(\omega^2 + i\,\omega/\tau_j)} \quad (18)$$

, where $\sigma_0$ is the DC conductivity and $\tau_j$ is the electron relaxation time in electrode j, given by $\tau_j = 1/(aT_j^2 + bT_j^3)$ with $a=10^7$ s$^{-1}$ K$^{-2}$ and $b=2 \times 10^6$ s$^{-1}$ K$^{-3}$ [37], where $T_j$ is the temperature of the corresponding electrode.

## 3. Results and Discussion

Here, we will first present the main results of this work in terms of the interplay between near-field radiative coupling and the space charge effect and the role of the interelectrode distance, at various input heat fluxes and operating voltages (subsection 3.1). Subsequently, we will discuss the effect of some important device parameters, namely emissivity and heat transfer coefficient, on the results (subsection 3.2). Finally, we will describe the effect of imperfect electron absorption at the collector (subsection 3.3).

## 3.1. The role of the interelectrode distance on TEC performance

The emitter and collector work functions of a TEC are crucial parameters as electron emission depends on them exponentially. To obtain high thermionic emission at practically achievable temperatures, the emitter work function should be low. On the other hand, to maximize efficiency, a large voltage difference between emitter and collector is required, which in turn requires the collector work function to be considerably lower than that of the emitter. However, if the collector work function is very low, significant back emission from the collector will occur, which lowers the efficiency. Given these requirements, a reasonable set of values for the emitter and collector work functions is $\varphi_E$=2.2 eV and $\varphi_C$=1.5 eV, which allows us to investigate the performance of the thermionic device for a wide range of other parameters. Moreover, these values of electrode work functions are also found to be optimal for the range of electrode temperatures considered in this study [18]. We use a Richardson constant of 120 $Acm^{-2}K^{-2}$, an effective emissivity of 0.1 for the radiation loss to the ambient from the emitter, and a heat transfer coefficient of 1000 $Wm^{-2}K^{-1}$ between the collector and cooling fluid (which is the upper bound of heat transfer coefficient for cooling by free convection). The choice of the aforementioned effective emissivity value was inspired by the designs of selective absorbers for concentrated solar power harvesting. Numerous design examples of such selective absorbers have been reported [38–42]. Such a material has a carefully designed spectral emissivity/absorptivity so that it can absorb most of the incident radiation spectrum from the sun (which is mostly around the visible range) and emit a relatively small amount of power through thermal radiation at a much lower temperature in a different spectral range (usually in the infrared range where the material has a very low emissivity). Nevertheless, the exact value of the effective emissivity could vary depending on the material properties of the absorber; we discuss the role of the effective emissivity of the emitter front side on TEC performance later. We assume the input heat source to have a power density of 10 $Wcm^{-2}$ unless specified otherwise. This power density is chosen to reach an emitter temperature that would result in a noticeable thermionic emission current. Such a power density can, for example, be obtained from a concentrated solar thermal harvesting system for a concentration factor of the order of 100 (assuming an AM 1.5 solar spectral irradiance). The solar thermal harvesting system can either employ selective absorbers or a liquid-parabolic trough concentrator mechanism to transfer the heat to the TEC emitter. It should be noted that we do not put any emphasis on the nature of the thermal source, and the presented model is equally valid for any source of input thermal energy. We vary the interelectrode distance from 0.1 µm to 100 µm to investigate the interplay between the space charge effect and near-field coupling in radiative heat transfer, while the output voltage is being swept.

Based on the energy balance at the emitter and collector as defined in eqs. (1) and (3), respectively, the emitter and collector temperature, net current density and different energy fluxes are numerically obtained using the self-consistent iterative method (shown in fig. 1(b)) for different operating voltages and interelectrode distances. Characteristic profiles of the emitter and collector temperatures as a function of the operating voltage, for an interelectrode distance of 5 µm, are illustrated in fig. 3(a). The corresponding current density and energy flux contributions are plotted in figs. 3(b) and 3(c), respectively. As can be seen in fig. 3(a), the emitter temperature remains constant in the saturation region ($V<V_S$) and it gradually rises in the space charge and retarding regions ($V>V_S$). This can be explained using the motive graph shown in fig. 3(b). In the saturation region, the maximum motive is equal to the emitter work function, which is the minimum energy required by the electron to be thermionically emitted from the emitter. Accordingly, all the emitted electrons can reach the collector as there is no potential barrier in the interelectrode space, and so the current does not change with the voltage. Due to this constant current, various energy fluxes (as shown in fig. 3(c)), which remove the input heat flux from the emitter, remain unchanged in the saturation region and prevent the temperature from increasing as the voltage is swept. On the other hand, in the space charge and retarding regions ($V>V_S$), the maximum motive is higher than the emitter work function and it gradually rises with the output voltage. Consequently, only a portion of the emitted electrons, which have sufficient energy to overcome the additional energy barrier in the interelectrode space, can reach the collector. As a result, both the emitter current and the energy taken by the thermionic electrons decrease as the voltage increases, leading to an increase in emitter temperature. As can be seen in fig. 3(c), this increase in emitter temperature will increase the radiative heat transfer from the emitter to the collector as well as the heat loss to the ambient. These various heat loss mechanisms start to dominate the energy flux from the emitter as the device is driven deep into the space charge and retarding regions. On the other hand, the collector temperature and the heat flux released to the heat sink from the collector decrease monotonically with the increase of the operating voltage, which can be explained as follows. In the saturation region, the electrical power output increases gradually with the operating voltage while the heat loss to the ambient remains constant due to the constant emitter temperature. As a result, the heat dissipated in the collector gradually decreases with the increasing voltage. In the space charge and retarding regions, the energy flux received by the collector decreases as output voltage increases, leading to a decrease in the collector temperature (although the reduced output power tends to counter this effect and lead to an increase in the collector temperature at very high input heat flux—not shown here).

To investigate the effects of interelectrode gap width on TEC operation, the emitter and collector temperatures are shown in figs. 4 (a) and (b), respectively for a wide range of output voltage and interelectrode distance. The corresponding current density for different interelectrode distances and output voltage are shown in fig. 4(c). It can be seen from fig. 4(c) that the interelectrode distance has a significant impact on the performance of the thermionic converter. Both the saturation current and open-circuit voltage are low when the interelectrode distance is in the submicron region. This finding is different from the previously reported results [29] where the thermionic current increases monotonically as the interelectrode distance is made smaller, and the open-circuit voltage remains constant irrespectively of the gap size. This fundamental difference is due to the fact that in ref. [29], the electrode temperatures were assumed to be constant as the gap size was varied. However, when the input energy is constant, the assumption of constant temperature no longer holds. When the gap is very small, the near-field radiative heat transfer between the electrodes becomes dominant and a significant portion of the input energy flux is transferred from the emitter to the collector through this mechanism. Since the total input energy is constant, the energy flux carried by the thermionic electrons is thus small at a very small electrode gap, meaning the current density is low, despite the mitigation of the space charge effect. As the interelectrode distance increases, the near-field effect of thermal radiation, which is caused by the coupling of evanescent waves between the two electrodes, gradually weakens. As a result, radiative heat transfer diminishes, and the energy carried by the thermionic electrons starts to dominate. An increased thermionic contribution to energy transfer between the electrodes requires more electrons to be thermally emitted from the emitter. Consequently, the emitter temperature and the thermionic current also increase with an increasing interelectrode distance as can be seen from figs. 4(a) and 4(c), respectively. It should be noted that such an increase in current with the interelectrode distance is not monotonic and, beyond a certain point, the current starts to fall off as the gap-induced space charge effect becomes significant. On the other hand, as seen from fig. 4(b), the collector temperature decreases with increasing the interelectrode distance due to weakening radiative coupling to the emitter. The trends versus operating voltage for a fixed interelectrode distance can be explained using the same reasoning as pointed out above in analyzing fig. 3.

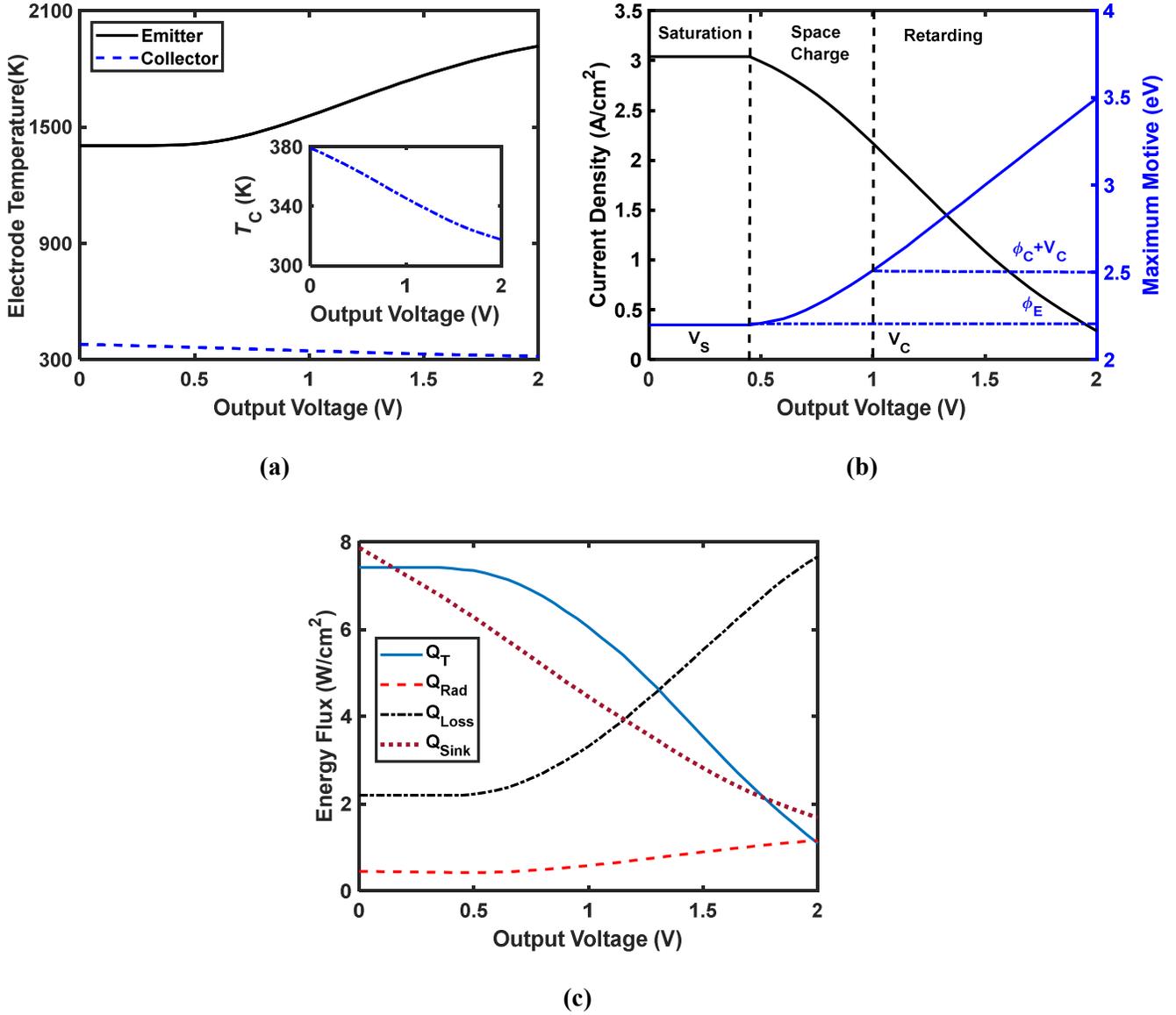

Fig. 3. (a) The electrode temperatures (the inset shows a magnified view of the collector temperature graph), (b) the current density (black curve) and the maximum motive (blue dashed curve), and (c) the energy fluxes from the emitter to collector and from the collector to heat sink as a function of the operating voltage for an interelectrode distance of 5 μm. The graphs are shown for an input heat flux of 10 Wcm$^{-2}$.

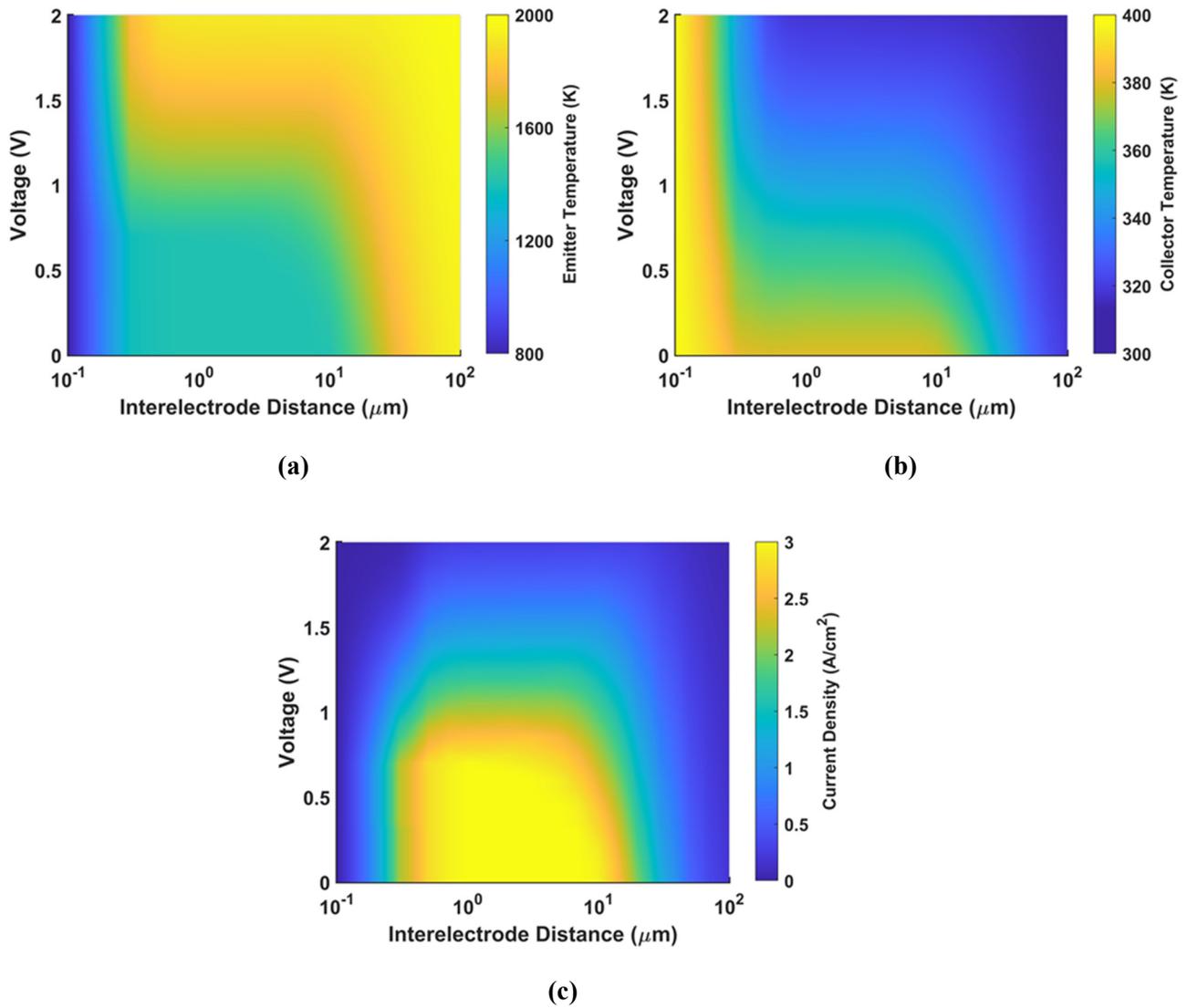

Fig. 4. Temperature graphs for (a) TEC emitter, (b) TEC collector, and (c) current density graph as a function of TEC output voltage and gap width. The graphs are shown for an input heat flux of 10 Wcm$^{-2}$.

Having understood the dependencies of the current and electrode temperatures on the interelectrode distance, we now turn to a detailed study of their implications on output electrical power, radiative heat transfer and device efficiency. Using eq. (2.a), (12), and (13), we calculate the output power density and interelectrode radiative heat transfer as a function of the operating voltage for a wide range of interelectrode distance; the results are shown in figs. 5(a) (where a maximum in TEC output power density with respect to both voltage and interelectrode gap width is seen) and 5(b), respectively. With the increase of interelectrode distance, the output power density initially increases as the near-field effect (which is a dominant radiative heat transfer process at very small gaps) weakens. However, this trend reaches a maximum for a certain interelectrode distance. Beyond this point, the space charge effect starts to dominate, which significantly reduces the electron flux from the emitter to the collector. As a result, the output power density starts to decrease again at large interelectrode distances, and radiative heat transfer (which is governed by the Stefan Boltzmann law at a large interelectrode distance) gradually becomes dominant again at a large interelectrode distance due to rising temperature difference between the electrodes.

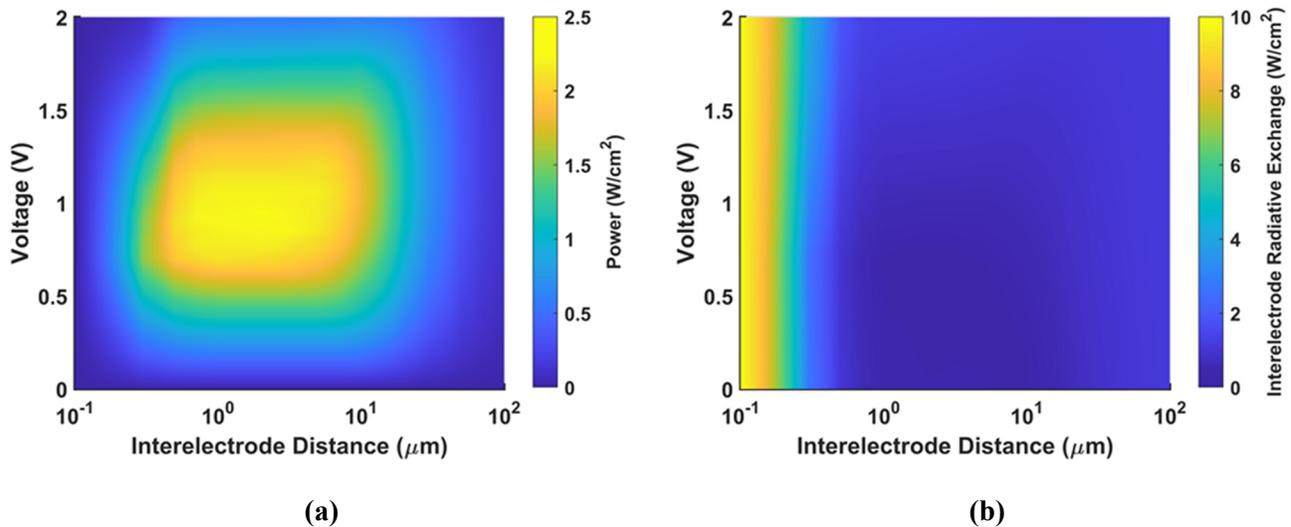

Fig. 5. (a) The output power density graph, and (b) interelectrode radiative exchange graph, as a function of TEC output voltage and gap width. The graphs are shown for an input heat flux of 10 Wcm$^{-2}$.

To better visualize the trends in device performance, the voltage-optimized conversion efficiency (which we also refer to as peak efficiency) is shown in fig. 6(a) as a function of interelectrode distance (together with the output voltage that corresponds to peak efficiency at each gap width); the corresponding emitter and collector temperatures are shown in fig. 6(b). The peak efficiency is low at small gaps due to the near-field effect and at large gaps due to the space charge effect. In-between these two regions, the peak efficiency reaches the maximum value of 22.28 % for d = 2 µm, which is of the order of the characteristic wavelength of thermal radiation (given by Wien's displacement law). The operating voltage corresponding to peak efficiency increases with the interelectrode distance as the device is gradually driven into the space charge region; the corresponding emitter temperature also increases while the collector temperature decreases with increasing distance.

To gain deeper insight, we now study the trends in the different energy fluxes under peak efficiency condition for a wide range of interelectrode distance values (0.1 µm to 100 µm) as shown in fig. 7. At very small distances, the energy exchange between the emitter and collector is dominated by near-field radiation, and the contribution from the thermionic exchange is very small. This is in stark contrast with the behaviour of the thermionic energy flux reported in ref. [29], where it remains constant as the interelectrode distance becomes very small. This is because, as we have already mentioned, in ref. [29] it was assumed that the heat source can provide any amount of power and the heat sink can remove any amount of unwanted heat required to maintain the emitter and collector temperatures constant irrespectively of the interelectrode distance. However, this assumption may not be valid in many practical scenarios. For example, if we consider the case of a solar-thermal-powered TEC, the energy input from solar radiation is limited by the atmospheric conditions, geographic location and concentration factor of the focusing system.

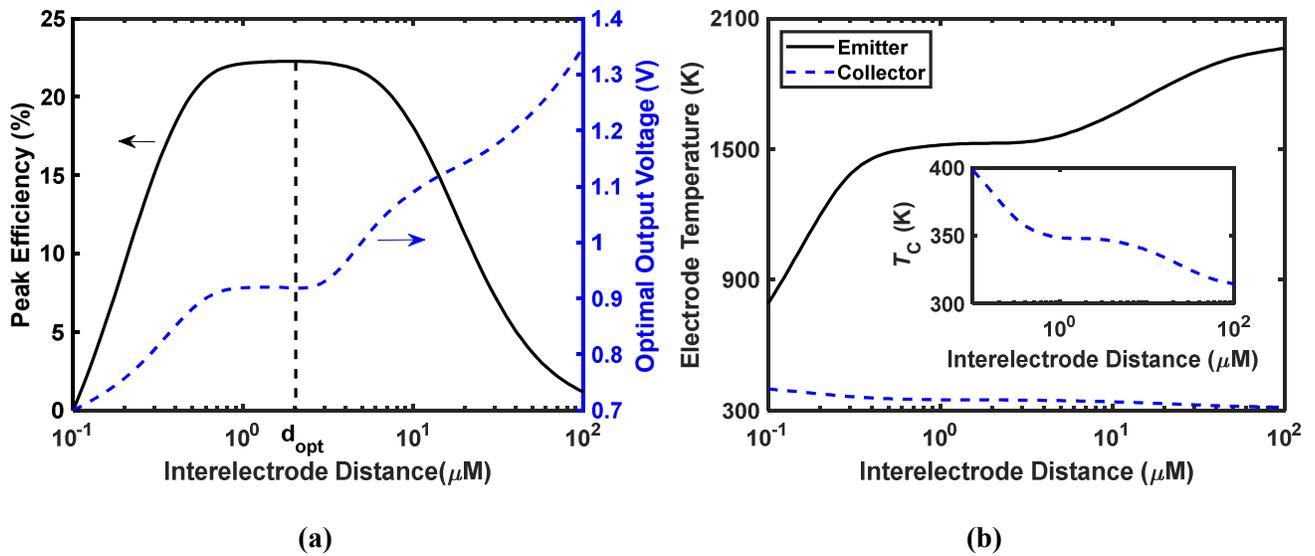

Fig. 6. (a) The peak efficiency and the corresponding optimal output voltage, and (b) the electrode temperatures at the optimal condition as a function of the interelectrode distance (the inset shows a magnified view of the collector temperature graph). The graphs are shown for an input heat flux of 10 Wcm$^{-2}$.

As well, due to the finite heat transfer coefficient between the collector and cooling mechanism, the collector temperature has to vary to accommodate different heat fluxes. This is why, in the present work, the emitter and collector temperatures are self-consistently determined using the energy balance condition.

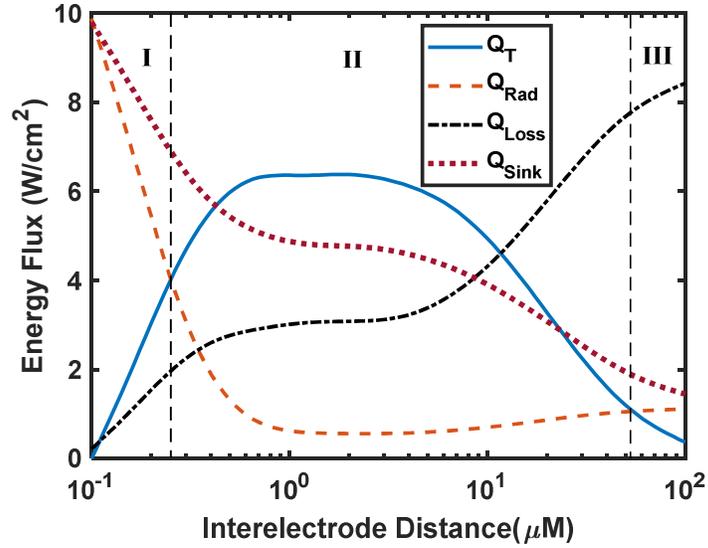

Fig. 7. The energy fluxes as a function of the interelectrode distance where the operating voltage has been optimized for maximum efficiency at each gap size. The interelectrode distances are divided into three regions as I) near-field dominated, II) thermionic dominated, and III) space charge dominated. The graphs are shown for an input heat flux of 10 Wcm$^{-2}$.

As the interelectrode distance increases, the radiative coupling becomes progressively weaker and thermionic coupling correspondingly stronger, so that the two cross over at a distance of around 0.25 µm. Meanwhile, the gradual increase in emitter temperature also necessitates an increase in radiative heat loss from the emitter to the ambient. Both the interelectrode thermionic and radiative exchange fluxes reach a plateau and this continues until space charge starts to affect the device performance, leading to a decrease in thermionic exchange and a corresponding increase in emitter temperature and thus radiative exchange (also in contrast with the behaviour observed in ref. [29]), resulting in another cross over at around 53.4 µm; the heat flux dissipated by the collector and released to the heat sink continues to decrease due to the decreasing total energy flux into the collector. Based on the above discussion, we define three regions of interelectrode distance as near-field radiation dominated (d < 0.25 µm), thermionic dominated (0.25 µm < d < 53.4 µm) and space charge dominated (d > 53.4 µm), respectively. (Note that the exact bounds of these regions depend on the material parameters such as permittivity, effective emissivity, and work function, but the above numbers provide an order-of-magnitude guideline.)

It is also worth investigating the device performance as a function of input energy. For example, in the case of solar thermal harvesting, the input heat flux can be varied by changing the concentration factor of the solar thermal receiver. (We note that, in a practical solar thermal receiving system, such wide-scale variations will not be possible once the system is physically implemented. However, when it comes to designing an optimal solar concentrator, one can always vary the concentration factor to observe its impact on TEC performance.) Fig. 8(a) shows that the maximum value of peak efficiency (maximized with respect to both output voltage and interelectrode distance) and the corresponding output voltage increase with input heat flux. The emitter and collector temperatures are shown in fig. 8(b), where we also observe a wide variation of these two electrode temperatures as a function of input heat flux under the optimal operating condition of the TEC. Since the TEC performance metrics have a strong nonlinear dependence on the electrode temperatures, these large temperature variations again emphasize the importance of considering the energy balance criterion when evaluating the conversion performance by the thermionic mechanism. For example, in fig. 8(a), note that the rate of increase in efficiency gradually slows down at higher heat fluxes. The reason is that the increased heat flux will raise the emitter temperature, which leads to more radiative heat loss from the emitter to the ambient and collector as well as an increase in space charge effect due to increased electron emission. Moreover, the resulting increase in collector temperature would affect energy balance at the emitter and also lead to significant back emission, which would work against an increase in efficiency, as also observed in ref. [43].

Inspired by ref. [29], we define a desirable gap range ($d_{min} < d < d_{max}$, where the efficiency is larger than 90% of $\eta_{max}$) and plot it as a function of input heat flux (fig. 9). An important observation is that the optimal gap size (corresponding to $\eta_{max}$), also shown in fig. 9, is relatively insensitive to input heat flux, which is a fortunate outcome from a device engineering point of view. We also see that the desirable range downshifts and becomes narrower with increasing input heat flux. This is in qualitative agreement with the results of ref. [29]. However, it should be noted that the lower limit of the desirable range and the optimal gap size found in this work are significantly higher than those found in ref. [29] for a comparable range of efficiencies (again due to the energy constraint issues discussed before). As an example, for efficiencies of 23% and 33%, the optimal gap sizes found in ref. [29] were around 900 nm and 800 nm, respectively, while, in the present work, they are more than twice higher (around 2 µm and 1.7 µm, respectively). Similarly, we find that the lower limits of the desirable gap range for the above-mentioned efficiencies are, respectively, 1.5 and 1.9 times higher than the values reported in ref. [29]. This is because, under the input heat flux constraint considered in this study, the lower bound of the desirable gap range reported in ref. [29] results in emitter temperatures that are too low to obtain any significant conversion efficiency. Although these differences may not appear significant at first glance, given the enormous materials and device fabrication related challenges (due to mechanical precision, surface roughness, and thermal expansion) associated with creating and maintaining microscale gaps between macroscopic electrodes in a wide range of temperatures, even a relaxation of the requirements by a factor of ~2 can be very significant for device engineering—another fortunate outcome of the present study.

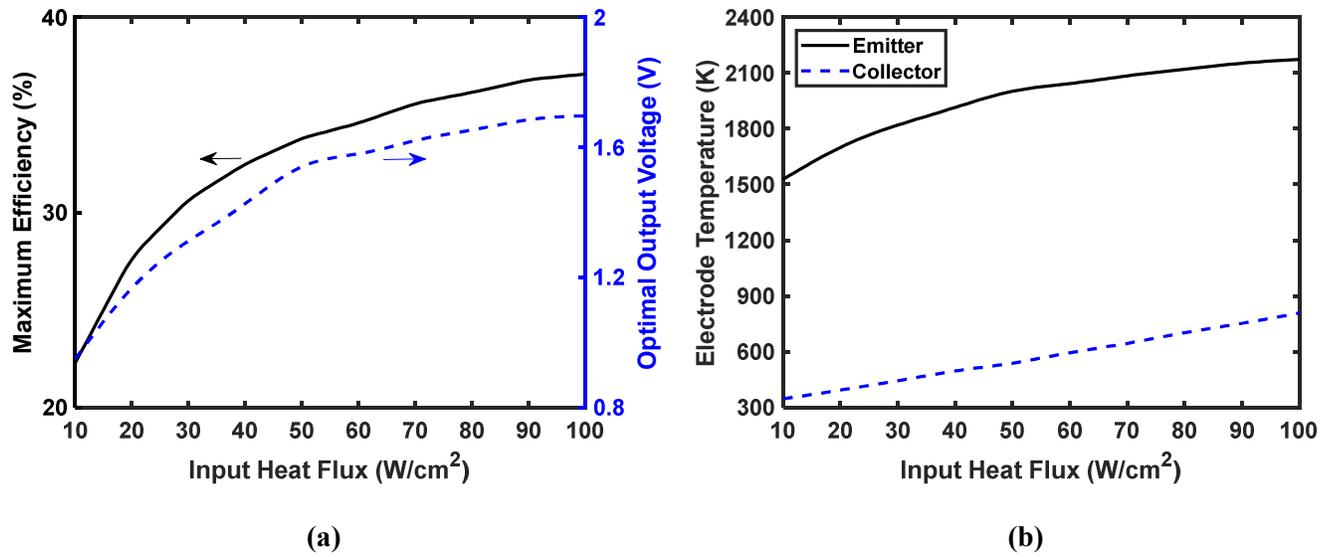

Fig. 8. (a) The TEC maximum efficiency (black curve) and the corresponding optimal operating voltage (blue dashed curve) as a function of the input heat flux. The curves shown are for the optimized interelectrode distance. (b) The emitter and collector temperatures as a function of the input heat flux where the gap size and operating voltage are optimized for maximum efficiency.

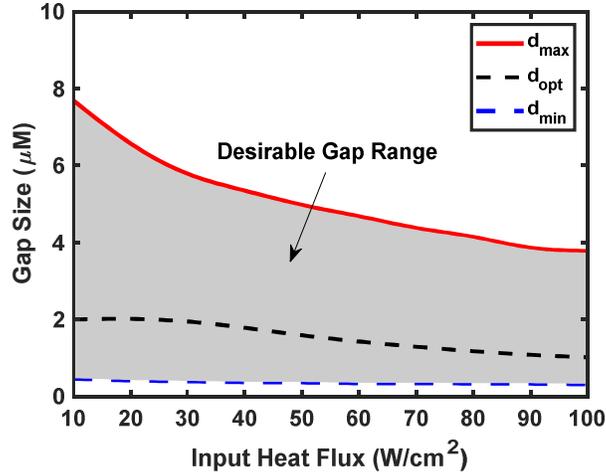

Fig. 9. Desirable gap range (shaded area) for different values of input heat flux.

### 3.2. The effect of the values of emissivity and heat transfer coefficient

It is also worth discussing the effect of key device parameters on the conversion performance. One such parameter is the effective emissivity of the emitter surface radiating thermal energy to the ambient. We have varied the effective emissivity over a wide range while keeping the other parameters at the values given before.

Fig. 10(a) shows that, as expected, the emitter and collector (which is radiatively coupled to the emitter) temperatures decrease monotonically with the increase of effective emissivity, because the latter leads to stronger radiative loss. The change in emitter temperature is much more noticeable compared to the collector because of the strong thermal contact of the collector to a heat sink. The maximum efficiency (the efficiency value optimized for both the TEC voltage and gap width) also decreases with a decrease in emitter temperature. On fig. 10(b), we see that the optimal gap width initially decreases with the increase of emissivity. We attribute this to the reduced electrode temperatures, which result in a weaker radiative exchange between the electrodes, thereby allowing a narrower gap width to mitigate the space charge effect further. After a certain point, the optimal gap width gradually increases with emissivity. This is because, as the emissivity increases, the gap size at which space charge sets in becomes larger. On the other hand, the optimal voltage gradually decreases with the increase of emissivity and approaches the flat band value of a space charge-free TEC.

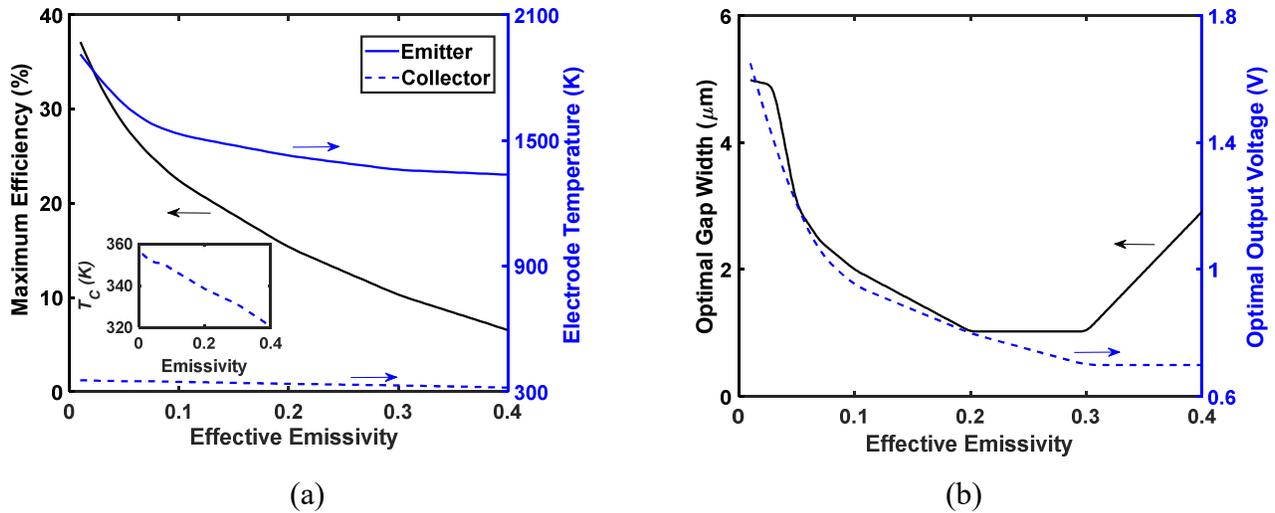

(a) (b)

Fig. 10. (a) The maximum efficiency and electrode temperatures as a function of the effective emissivity (over the spectral range where it radiates thermal energy to the ambient) of the emitter. The inset shows a magnified view of the collector temperature graph. (b) The optimal gap width and output voltage as a function of the effective emissivity of the emitter. The graphs are shown for an input heat flux of 10 Wcm$^{-2}$.

It is also interesting to analyze the effect of the collector heat transfer coefficient on device performance (fig. 11). (We emphasize that a heat transfer coefficient higher than $10^3$ Wm$^{-2}$ K$^{-1}$ (which we have used so far) would require forced convection or need a phase change mechanism [44,45]. Therefore, additional power would be required to pump the cooling fluid to remove the heat from the collector, which we neglect in the efficiency analysis for simplicity (a common practice in the literature).) The $K_L$ value of $10^4$ Wm$^{-2}$ K$^{-1}$ was chosen as it was found to keep the collector temperature very close to the ambient over a wide range of input heat flux. The other parameters are kept at the values used before.

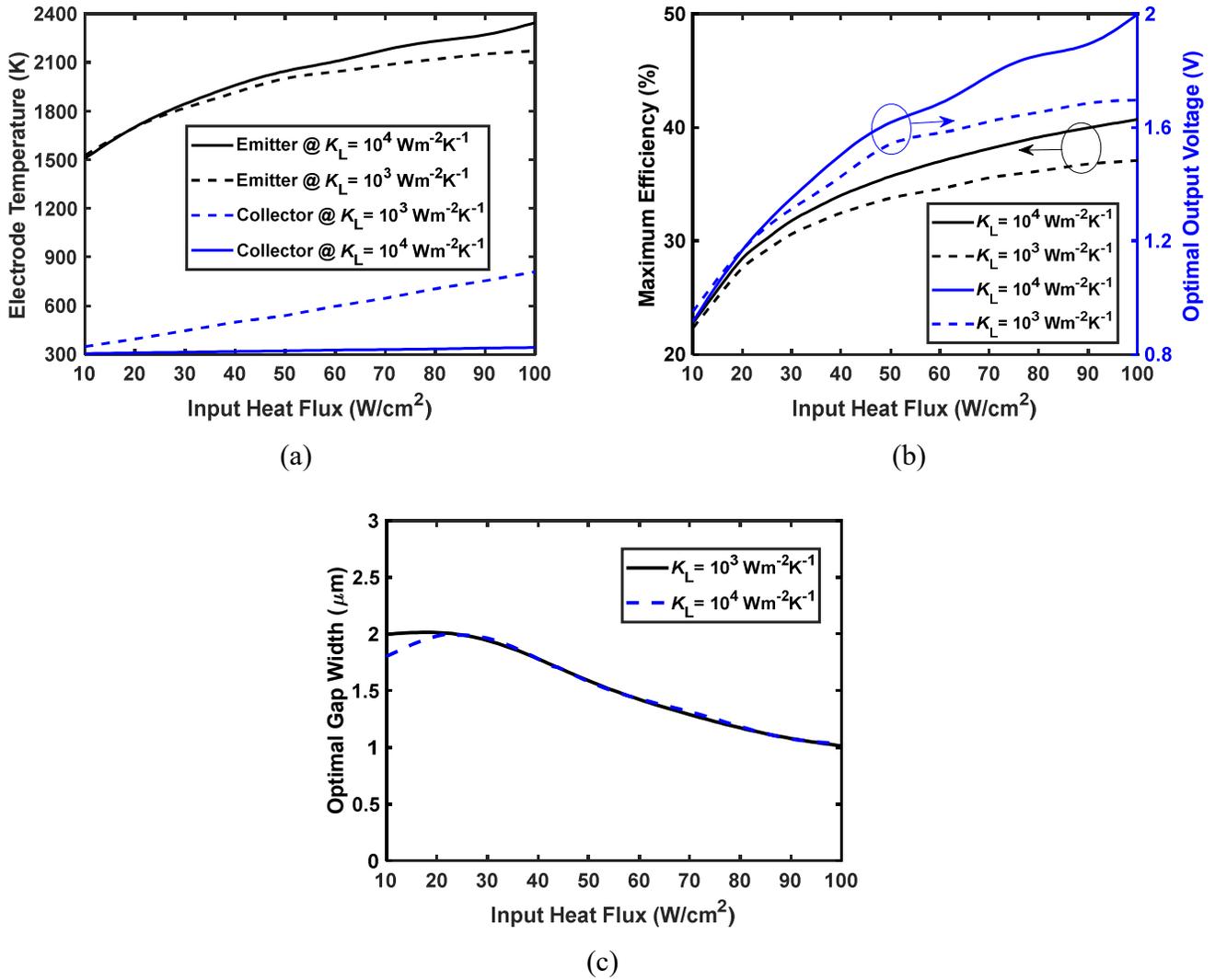

Fig. 11. (a) The electrode temperatures, (b) maximum efficiency and the corresponding optimal operating voltage, and (c) optimal gap width as a function of the input heat flux for different heat transfer coefficients.

As can be seen from fig. 11(a), the collector temperature decreases when we change $K_L$ to $10^4$ $\text{Wm}^{-2}\text{K}^{-1}$; however, the emitter temperature increases in most of the range of the input heat flux values studied. The latter is because the dielectric properties of tungsten are such that, in this temperature range, a decrease in the collector temperature leads to weaker near-field radiative energy exchange. The increase in emitter temperature and decrease in collector temperature naturally lead to an increase in the maximum conversion efficiency (fig. 11(b)). In addition, the higher TEC current also leads to higher space charge and thus an increase in the optimal output voltage (fig. 11(b)). In fig. 11(c), we see that, for relatively low input heat flux, the optimal gap width is slightly reduced for the higher heat transfer coefficient. We attribute this to a reduced radiative coupling (as explained before) which allows the electrodes to be placed slightly closer to further mitigate the space charge effect. Interestingly, at higher heat flux values, the contradictory requirements of reducing the space charge effect and interelectrode radiative exchange result in an optimal gap width which is very similar for both $K_L$ values.

### 3.3. Operation under imperfect electron absorption by the collector

Finally, we discuss the effect of electron reflection from the collector. To study this issue, we have added the physics of electron reflection from the collector in our self-consistent device model. The relevant theoretical details can be found in [46] and will not be repeated here.

Here, we will show the key findings of this additional study of collector reflection. For this purpose, we assumed that the collector is 50% reflective, which is the average between the worst case (completely reflective) and best case scenario (perfectly absorbing) and compared it with the perfectly absorbing case. The other device- and material-related parameters are kept identical to the values as mentioned at the beginning of the "Result and Discussion" section to solely investigate the effect of electron reflection from the collector.

In fig. 12(a), we show the current density as a function of the output voltage for different interelectrode distances. The gap width values are chosen so that the transition from the emission limited to space charge limited regime can be clearly seen. We found that, for a reflective collector, the space charge effect sets in at a lower gap width, and it is observed over a wider output voltage range. This is expected because a higher density of electrons is present in interelectrode space due to electron reflection from the collector. Moreover, the effect of this more dominant space charge due to collector reflection can be easily seen on the device performance as shown in fig. 12(b). The maximum value of the peak efficiency and the optimal gap width decreases in the presence of collector reflection. In addition, the desirable gap range becomes narrower.

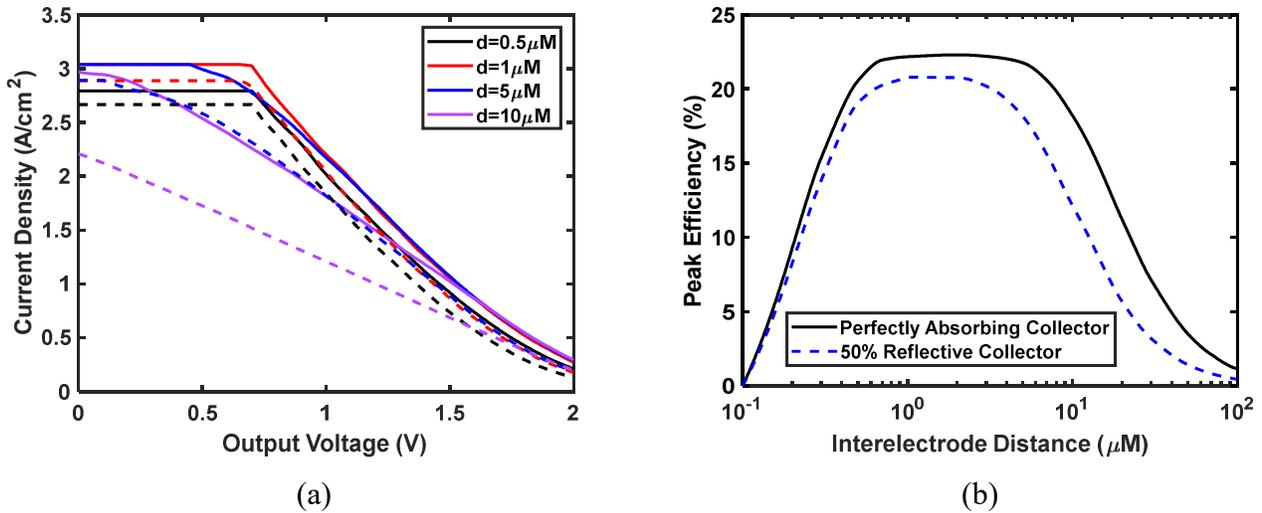

Fig. 12. (a) The current-voltage characteristics for different interelectrode distances. The solid lines represent the results for a perfectly absorbing collector and the dotted lines are for a 50% reflective collector. (b) The peak efficiency curves as a function of the interelectrode distance for a perfectly absorbing collector and a 50% reflective collector. The graphs are shown for an input heat flux of 10 Wcm$^{-2}$.

## 4. Summary

In summary, we presented a comprehensive analysis of the operation of a micro gap vacuum thermionic energy converter under the constraint of a fixed input heat flux. Space charge and near-field coupling of thermal radiation, which are key effects in the energy exchange processes, were taken into consideration. The input energy constraint considered here requires the device performance to be analyzed under a numerical, iterative framework based on the energy balance at each electrode. The results elucidate the dependencies of the electrode temperatures, various interelectrode energy fluxes, device current, and output power density on interelectrode distance. An efficiency optimization in terms of operating voltage and interelectrode distance was also carried out, showing that the desired interelectrode distance can be considerably higher than previously thought. The influence of the input heat flux on electrode temperatures, efficiency, and operating voltage was also discussed. The effects of the emitter effective emissivity, collector cooling capacity, and electron reflection from the collector on device performance were also illustrated with various examples. The findings of this work thus provide useful insights into the device physics of a micro gap TEC which are crucial for the design and fabrication of thermionic energy harvesting systems.

## Acknowledgement

We acknowledge financial support from the Natural Sciences and Engineering Research Council of Canada (RGPIN-2017-04608, RGPAS-2017-507958, SPG-P 478867). This research was conducted, in part, by funding from the Canada First Research Excellence Fund, Quantum Materials and Future Technologies Program. Ehsanur Rahman thanks the Natural Sciences and Engineering Research Council of Canada for a Vanier Canada Graduate Scholarship and the University of British Columbia for an International Doctoral Fellowship and Faculty of Applied Science Graduate Award.## Author Contribution

Ehsanur Rahman conceptualized the work, developed algorithm, performed simulation, model verification, data visualization, and wrote the draft manuscript; Alireza Nojeh provided technical guidance, reviewed, and edited the manuscript.

## References

[1]   L. C. Hirst and N. J. Ekins-Daukes, Fundamental losses in solar cells, Prog. Photovolt. **19**, 286 (2011).

[2]   C. H. Henry, Limiting efficiencies of ideal single and multiple energy gap terrestrial solar cells, J. Appl. Phys. **51**, 4494 (1980).

[3]   D. J. Friedman, Progress and challenges for next-generation high-efficiency multijunction solar cells, Curr. Opin. Solid State Mater. Sci. **14**, 131 (2010).

[4]   C. O'Dwyer, R. Chen, J.-H. He, J. Lee, and K. M. Razeeb, Scientific and Technical Challenges in Thermal Transport and Thermoelectric Materials and Devices, ECS J. Solid State Sci. Technol. **6**, N3058 (2017).

[5]   W. Liu, Q. Jie, H. S. Kim, and Z. Ren, Current progress and future challenges in thermoelectric power generation: From materials to devices, Acta Mater. **87**, 357 (2015).

[6]   A. H. Khoshaman, H. D. E. Fan, A. T. Koch, G. A. Sawatzky, and A. Nojeh, Thermionics, thermoelectrics, and nanotechnology: New possibilities for old ideas, IEEE Nanotechnol. Mag. **8**, 4 (2014).

[7]   D. B. Go, J. R. Haase, J. George, J. Mannhart, R. Wanke, A. Nojeh, and R. Nemanich, Thermionic Energy Conversion in the Twenty-first Century: Advances and Opportunities for Space and Terrestrial Applications, Front. Mech. Eng. **3** (2017).30